\documentstyle[aps,epsfig,multicol,graphicx,prl]{revtex}

\begin{document}

\title{Thermal convection in fluidized granular systems\thanks{URL
http://www.cec.uchile.cl/cinetica/ }}
 
\author{ Rosa\, Ram\'{\i}rez\footnote{Present address: 
 CECAM, ENS-Lyon, 46 All\'ee d'Italie, 69007, France}$,^{1}$  Dino\, Risso$,^{2}$\,
and \, Patricio\, Cordero$^{1}$ \\ {(1)\,~Departamento de
F\'{\i}sica, Facultad de Ciencias F\'{\i}sicas y Matem\'aticas
\\ Universidad de Chile, Santiago, Chile \\ (2)\, Universidad
del B{\'{\i}}o B{\'{\i}}o, Departamento de F{\'{\i}}sica, Concepci\'on,
Chile} }

\maketitle 

\begin{abstract}

Thermal convection is observed in molecular dynamic simulation of a
fluidized granular system of nearly elastic hard disks moving
under gravity, inside a rectangular box.
Boundaries introduce no shearing or time dependence, but the energy
injection comes from a slip (shear-free) thermalizing base. The
top wall is perfectly elastic and lateral boundaries are either
elastic or periodic. The observed convection comes from the
effect of gravity and the spontaneous granular temperature
gradient that the system dynamically develops.


\end{abstract}

\begin{multicols}{2}

In the study of granular systems, convection has attracted
particular attention. Many of the experimental and simulation
studies of granular convection focus their attention on the
effects of a vibrating base~\cite{herrmann92,aoki96,ehrichs95,knight96,
 pak95,saluena98,tennakoon98b,grossman97}.  Particular care has been taken
to determine the different patterns that appear depending on the
amplitude and frequency of the vibrating base, the form of the
interaction force between the grains, the effects caused by the
roughness and inclination of the walls and the role of voids in
this vibrating-base-plus-wall convection. Some theories based on
hydrodynamic continuum equations for this type of convection
have been
developed~\cite{bourzutschky95,hayakawa95b,hayakawa97b}. Also
the effect of the internal shear bands in the system has been
pointed out as a source of convection~\cite{knight97}.

The purpose of this letter is to point out the existence of a
convective regime in granular systems when there is no vibrating
base, or any effect caused by the roughness or the angle of the
walls, or any shearing effect caused by the walls
themselves.  The source of this convection stems from {\em
gravity} and the dissipative nature of the granular collisions, and some
hint about its existence was already mentioned in \cite{bizon98}.
In the present model system, each wall is represented by a
shear-free and time-independent boundary condition.

More precisely the bottom wall gives 
a stochastic normal component to the
velocity of each particle that collides with it.
The tangential velocity is unchanged, thus a shear-free thermal boundary
condition is imposed. The top wall is perfectly elastic while the lateral
walls are either perfectly elastic or they correspond to a
periodic boundary. Particles are subjected to the acceleration
of gravity $g$.

When energy is pumped into a granular system through a
boundary---the bottom wall in the present case---a temperature
gradient develops spontaneously. Since the particle-particle
collisions are dissipative the system is hotter close to the
injecting boundary and colder away from it. There is then an
energy flux from the base upwards which is dissipated
in the bulk through collisions. If the system is almost
perfectly elastic it remains macroscopically static.

Performing molecular dynamic simulations of inelastic hard disk
systems in the presence of gravity, a transition from a
hydrostatic (purely conductive regime) to a convective regime is
observed by increasing the dissipative parameter present in the
particle-particle collision rule.

In standard fluids the onset of  thermal convection is roughly
determined by the ratio between the characteristic times of the
processes against convection (viscous and thermal
diffusion) and favorable to convection (buoyancy), through
which the externally imposed temperature gradient plays a
role~\cite{RB}.  The idea that underlies this letter is that in
a granular system there are similar mechanisms which lead to a
transition from a conductive to a convective regime.

There is an important difference, though. The temperature
gradient is not externally imposed but rather it is dynamically
created by the system itself. This means that whenever energy is
injected into the system a temperature gradient develops (due to
the dissipative nature of the particle-particle collisions) and
not three but four ingredients compete inhibiting or fostering
the appearance of convection: viscous and thermal diffusion,
buoyancy {\em and dissipation}.

This letter reports  the results about convection appearing in
a 2D system of $N$ hard disks with mass $m=1$ and diameter
$\sigma=1$, which collide inelastically with the rule
\begin{equation}
\vec{v_{12}}^{\,\prime}\cdot \hat{n}= - (1 - 2 q)
\,\left(\vec{v_{12}}\cdot\hat{n}\right)\,, \qquad
\vec{v_{12}}^{\,\prime}\cdot \hat{t}=
\,\vec{v_{12}}\cdot\hat{t}\,,
\label{eq:colision}
\end{equation}
where $\vec{v_{12}}=\vec{c}_1-\vec{c}_2$ is the relative velocity
between the colliding particles, the primed and unprimed
variables refer to the post and precollisional velocities,
$\hat{n}$ and $\hat{t}$ are the unit vectors normal and 
tangential to the contact plane, and $q$ is the dissipative
coefficient $q=(1-r)/2$, $r$ being the normal restitution
coefficient.  Only translational degrees of freedom are present.
The simulation results reported below come from our
event-driven molecular dynamic simulations~\cite{MRC},
and the careful measuring routines developed in~\cite{tesisdino}.

As already mentioned, the system is maintained in a fluidized
state by the injection of energy from a thermal (stochastic)
base ($y=0$) at temperature $T(0)=T_{\rm base}=1$ in energy units,
while the top boundary is a perfectly elastic wall at $y=L$. 
Gravity enters the problem through the Froude number,
$$F\!r =\frac{mgL}{T_{\rm base}}\, . $$

Although there is a wide range of parameters for which
convection occurs, all the simulations being reported correspond
to $N=2300$ particles, fraction of occupied area $\rho_A=0.18$,
and relative gravity strength $F\!r = 0.1$. The only parameter
that varies from one simulation to another is $q$, even though
for convenience the product $qN$ is used in what follows.

\bigskip

\begin{figure}[htb]
\begin{minipage}{0.9\hsize} 
\noindent
\begin{flushleft}
\centerline{\includegraphics[width=1.2\textwidth,angle=180]{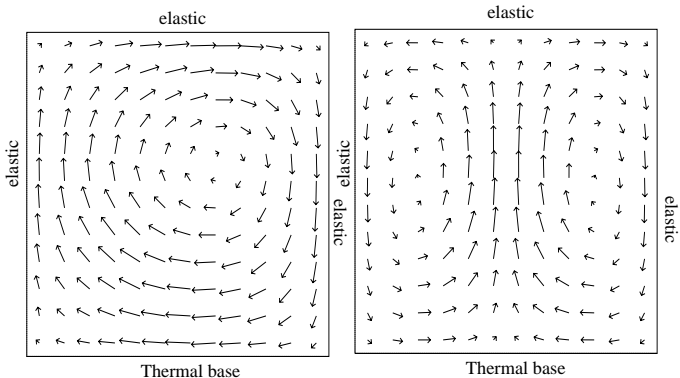}}
\end{flushleft}
\bigskip
\centering
\caption{\protect \label{figure1} 
Averaged velocity field for different values of $qN$. At the left  
$qN=6$ and at the right $qN=34$.}
\end{minipage}
\end{figure}

\paragraph{Convection inside a box.} \ Consider first the
granular system inside a squared box with a thermal base, and perfectly
elastic upper and lateral walls.  As the elastic walls do not
produce any direct shear on this system and the bottom base is
stochastic but preserving the horizontal component of the
velocity, the boundaries introduce neither spatial nor temporal
macroscopic correlations. It can be said that none of the usual
conditions under which convection has been studied in granular
systems are present, nevertheless convection does appear, as observed in
Fig. \ref{figure1}.

The hydrodynamic stationary solution for the associated conservative
system ($q=0$) is simply a constant temperature system with no
heat flux and density decreasing with height. Including a small amount of
energy dissipation in each collision, one and even multiple rolls may develop 
into the system. Hence the convection we observe 
is due solely to gravity and the
dissipative nature of the collisions between particles.

\begin{figure}[htb]
\begin{minipage}{0.9\hsize} \centering \noindent
\includegraphics[width=.65\textwidth,angle=270]{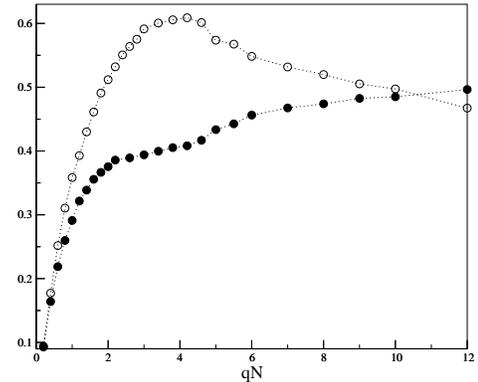}
\caption{\protect \label{figure2} Differences between the bottom
and top temperatures (solid circles) and rescaled heat flux
(open circles) versus $qN$}

\end{minipage}
\end{figure}

To understand the origin of this convection, 
we have plotted (see Fig.\ref{figure2}) the dependence on the dissipative parameter $qN$ of the
observed granular temperature difference between the bottom ($y=0$) and the
top of the system ($y=L$), $\Delta=T(0)-T(L)$.  For small values of $qN$ this
difference increases with increasing $qN$.  This situation corresponds to
the conductive regime, in which $\Delta$ can be written as an expansion on
the small parameter $qN\rho_A$~\cite{tesisrosa}. This difference $\Delta$
reaches a maximum, at about $qN\approx 4$, and from then on it decreases,
proving that a mechanism favoring energy transport
from the base upwards appears at this value. This fact is also corroborated by
the amount of energy per unit time (call it heat flux) $Q(0)$, entering the
system through the base, plotted in the same figure in arbitrary units. It
is seen that the heat flux is steeper about the same $qN$ for which $\Delta$
reaches its maximum: more energy per unit time is required to keep stable
the temperature at the base.
  
From these observations we can conclude that 
there exists a threshold $qN$ value from which
the convection is triggered. Furthermore, this also supports the idea that
convection starts when a critical value of the 
temperature difference between the bottom and the top of the system is
reached. 

\vspace{0.5cm}

\begin{figure}[htb]
\begin{minipage}{0.9\hsize} 
\centering 
\noindent 
\includegraphics[width=0.7\textwidth,angle=270]{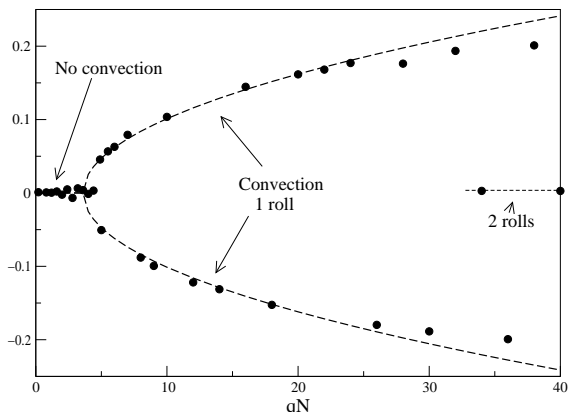} 
\caption{\protect \label{figure3} 
Mass circulation $\Phi$ measured in simulations (points). The dashed lines
correspond to the curve $\Phi = \pm 0.04 \sqrt{qN-3.8}$.} 
\end{minipage}
\end{figure}

A way to detect and quantify this transition is by measuring mass
circulation in the system. This is implemented by calculating the sum of
integrals of the velocity field along many concentric paths centered about
the geometric center of the box: $\Phi =\sum\int \vec v\cdot d\vec l$.  This
observable $\Phi$ will be negligible if there is no convection and it will
be distinctly nonzero (positive or negative) if there is one (anti)
clockwise convective roll. Observed values of $\Phi$ are plotted in Fig.
\ref{figure3} clearly showing a supercritical transition at $qN\approx4$
from the conductive to the convective regime with one convection roll. Due
to the symmetry of the problem, rolls with both signatures are equally
probable and they appear in our simulations, depending only on the initial
condition.

It can also be seen that from $qN \approx 34$ up there is a coexistence of
regimes with zero and non zero circulation which corresponds to the
competition of one and two convective rolls, this is, a subcritical
transition from the one-roll to the two-rolls regime. As the dissipation
coefficient $q$ continues increasing, other transitions to multi-rolls
patterns can be observed with difficulty but since the system starts getting
denser all convective movement eventually disappears.

\vspace{0.6cm}

Although no theory has been presented yet, it seems that the appearance of
the two and multi-rolls regimes could be due to a change in the effective
aspect ratio of the system: as dissipation increases, there are regions
where density rises considerably, lowering the average height occupied by
the system.

It is worth mentioning that when two rolls were observed they always
appeared as shown in Fig. \ref{figure1}, namely, the fluid goes up in the
middle of the box and comes down along the walls. This privileged signature
seems to have its origin in the local increase of density that walls cause.
Higher density implies more collisions and therefore more dissipation, hence
lower granular temperature: the system is heavier near the walls.

\bigskip

\paragraph{Convection with horizontal periodic boundary conditions.} \ Any
effect that the elastic lateral walls could have on the onset of convection
in the previous case is discarded when periodic lateral boundary conditions
are imposed on the system. 

The container is a periodic channel, and in this case a transition to a
convective regime is found again, although, due to the absence of lateral
boundaries, the convective rolls appearing in the system travel now along
the channel. This was observed even though the simulations were carefully
initialized with zero total horizontal momentum $P_x$, namely with zero
horizontal mass flux. Since the boundary conditions do not change the
horizontal component of the velocities, $P_x$ remains zero during the
evolution, as was confirmed in the simulation.

In order to detect this pattern, we performed time averages of the mass flux
field. Because of the roll movement, this averaging time must be larger than
the microscopic time and smaller than the time needed for the roll to travel
a significant distance. We chose this time to be much smaller than the ideal
gas thermal diffusion time which, in our units, is of order
$N\sqrt{\pi/T_{\rm base}}$, but large enough to contain multiple
particle-particle collisions. The observed rolls persisted for times longer
than the macroscopic time, resulting in an hydrodynamic pattern.

\begin{figure}[htb]
\begin{minipage}{0.9\hsize}
\noindent 
\begin{flushleft}
\centerline{\includegraphics[width=1.1\textwidth,angle=180]{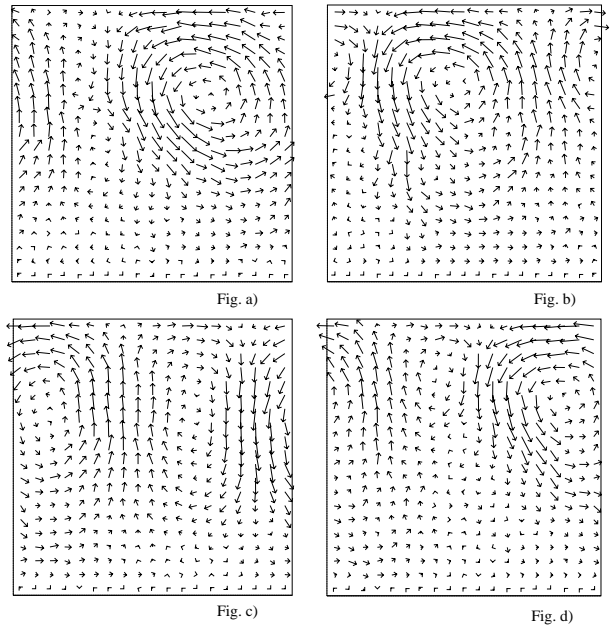}}
\end{flushleft}
\bigskip
\centering
\caption{\protect \label{figure4} Mass flux field averaged in circles of 
250 collisions per particle. The figures $a)$, $b)$, $c)$ and $d)$
correspond to cycles $100$,$112$,$124$ and $136$ respectively. A big roll
can be observed moving to the left side of the system while a small roll
appears varying its size.}     
\end{minipage}
 
\end{figure}

An example of what is happening in the system can be observed in figure
\ref{figure4}. This figure is a plot of the averaged mass flux field at four
different stages of the simulation.  Due to the periodic lateral boundaries,
the solution should be a two-rolls pattern, (or any even number of rolls)
but the aspect ratio forced on the system would imply rolls with a width
about half their height, which makes them unstable.  The system was most of
the time observed to have one large roughly circular roll accompanied by a
smaller one.

The reason for this movement seems to be found in vortex dynamics. It is
well known that a vortex near a fixed isolating wall behaves as if it were
in front of a twin vortex which ensures the condition of null hydrodynamic
velocity $v_y=0$ at the top elastic wall~\cite{saffman}.  This twin vortex
would induce a movement parallel to the wall and a sense determined by 
the sign of the circulation of the original vortex.

\vspace{0.4cm}

\begin{figure}[htb]
\begin{minipage}{0.9\hsize}
\centering 
\noindent 
\includegraphics[width=.7\textwidth,angle=270]{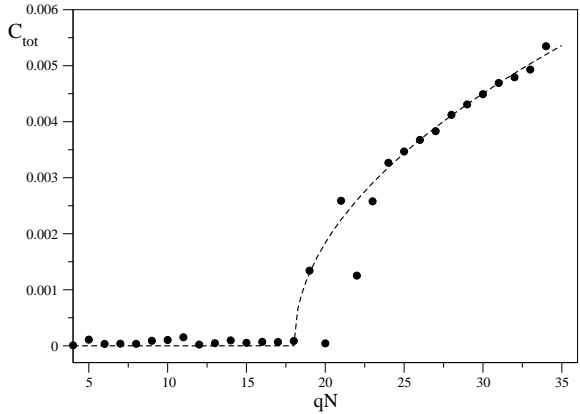}
\medskip
\caption{\protect \label{figure5} Total velocity correlation $C_{\rm tot}$ vs. $qN$. The transition from
a conductive to a convective regime is observed at $qN \approx 18$. The data
have some dispersion near the transition due to the finite size of the
system. The dashed line corresponds to the curve $C_{\rm tot} = 0.0013
\sqrt{qN-18}$.}
\end{minipage}
 
\end{figure}

In the present case, to detect the onset of convection, it was not practical
to measure the circulation $\Phi$ as when the system was between hard
lateral walls. Instead the transition was detected measuring a space
velocity correlation. The system is tiled with cells $(i,j)$ and a
hydrodynamic velocity correlation is defined by
\begin{equation}
\small C(i,j)=\frac{1}{8} \sum_{i',j'} \vec v(i,j) \cdot \vec
v(i',j')\,,
\end{equation}
where the cells $(i',j')$ refer to the eight first neighbors of the $(i,j)$
cell. This observable has the advantage of being insensitive to the
displacements of the convective pattern.  The total correlation is defined
as $C_{\rm tot} = 1/N_{\rm cells} \sum_{i,j}C(i,j)$. It measures how
similar, on the average, are the velocities in neighboring cells. When there
is no convective pattern in the system, then the time averaged value of
$C_{\rm tot}$ is nearly zero, while as soon as a convective current
develops, $C_{\rm tot}$ takes distinctly positive value.

Figure \ref{figure5} shows the evolution of  $C_{\rm tot}$  with 
$qN$. The transition from conductive to  convective
regime is clearly seen. It takes place  roughly at $qN=18$.

\bigskip

In conclusion it can be stated that bidimensional granular
systems present convective regimes when there is gravity and a
thermal base even though no shearing is introduced through the
boundary conditions. It has been argued that such convection
owes its existence only to gravity and the dissipative nature of
the particle-particle collisions. If the system approaches the
elastic limit such convection disappears as there is no spontaneous
creation of a vertical temperature gradient. This gradient plays  
the key role in this kind of convection.

\bigskip

The authors wish to thank Prof. Rodrigo Soto for helpful
comments and discussions.
This work has been partly financed by Fondecyt research grant
296-0021 (R.R.), Fondecyt research grant 1990148 (D.R.), and
Fondecyt research grant 1970786 (P.C.).

\end{multicols}

\end{document}